\documentstyle[12pt,psfig]{article}

\def\e{\epsilon}

\def\k{\kappa}
\def\a{\alpha}
\def\b{\beta}

\def\s{\sigma}
\def\r{\rho}
\def\d{\delta}

\def\p{\phi_a}
\def\ptil{\tilde\phi_a}

\begin{document}

\begin{titlepage}
 \hfill       OHSTPY-HEP-T-99-020\\
  \mbox{ } \hfill      November 1999 \\

\vspace{1.0cm}
 
\begin{center}

  {\Large\bf Fermion Masses and
      Neutrino Oscillations}

\vspace{.5cm}

  {\Large\bf in $SO(10)$ SUSY GUT with $D_3 \times U(1)$}

\vspace{.5cm}

  {\Large\bf Family Symmetry}

  \vspace{2cm}

  {\large\bf              
    Radovan Derm\' \i \v sek and Stuart Raby}

    \bigskip
{\em Department of Physics, The Ohio State University, 174 W. 18th Ave.,
Columbus, Ohio  43210}

  \vspace{1cm}            
{\bf Abstract}
\end{center}

Discrete nonabelian gauge symmetries appear to be the most advantageous
candidates for a family symmetry. We present a predictive $SO(10)$ SUSY GUT
model with $D_3 \times U(1)$ family symmetry ($D_3$ is the dihedral group 
of order 6).  The hierarchy in fermion masses is generated by the family 
symmetry breaking $D_3 \times U(1) \rightarrow Z_N \rightarrow $  nothing. 
This model fits the low energy data in the charged fermion sector quite well
and naturally provides large angle $\nu_{\mu}$~--~$\nu_{\tau}$ mixing 
describing atmospheric neutrino oscillation data and small 
angle $\nu_e$~--~$\nu_s$ mixing consistent with the small mixing angle MSW 
solution to the solar neutrino data.
In addition,  the non-abelian family symmetry $D_3$ is sufficient to 
suppress large flavor violations.

\end{titlepage}

\section{Introduction}

The origin of the fermion mass hierarchy is one of the most challenging 
problems in elementary particle physics. In the standard model 
fermion masses and mixing angles are free parameters. Even though these 
13 parameters (9 charged fermion masses; 3 angles and 1 phase in the CKM 
matrix) are well known experimentally, the standard model does not offer 
any explanation. Supersymmetric [SUSY] Grand Unified Theories [GUTs] besides 
gauge coupling unification also provide relations between quark 
and lepton masses within generations.  However, the understanding of the 
hierarchy between generations is still missing. A possible solution to 
the fermion mass hierarchy problem is to introduce a new symmetry -- 
family symmetry -- acting horizontally between generations. The hierarchy 
is then generated by sequential spontaneous breaking of this symmetry. 
Furthermore, acting differently on different generations, family 
symmetries can provide a solution to the problem of large flavor changing 
neutral currents [FCNCs] in SUSY~\cite{flavorviolation}.

A variety of models~\cite{familysymmetry} --~\cite{tetrahedral} 
with family symmetries were proposed. Among these, models with $U(2)$ 
(or its subgroups) family symmetry~\cite{u2symmetry} -- ~\cite{tetrahedral} 
appear to be very promising candidates for the theory of flavor. The reason 
for this is twofold: the top quark is the only fermion with mass of order the 
weak scale, thus distinguishing the third generation from the others; and by 
placing the first and second generations into a two dimensional 
irreducible representation of the family group the degeneracy of squarks 
in these two generations can be achieved, which is necessary to suppress 
 FCNCs. Thus non-Abelian family symmetries, especially $U(2)$ or its 
subgroups, are naturally suggested.

We would like to focus here on a particular model presented 
in~\cite{BRT}.  It is an $SO(10)$ SUSY GUT with family symmetry $U(2) 
\times U(1)$.\footnote{The model~\cite{BRT} is a modification of the 
$SO(10) \times U(2)$ model suggested in~\cite{so10u2}. 
The modification only affects the results in the neutrino sector.} 
This model is ``predictive" by which we mean that it is 
``natural" -- the Lagrangian contains all terms consistent with the 
symmetries and particle content of the theory; and the number of 
arbitrary parameters is less than the number of observables. 
This model fits the low energy data in the charged fermion sector quite well
and naturally provides large angle $\nu_{\mu}$~--~$\nu_{\tau}$ 
mixing describing atmospheric neutrino oscillation data and small 
angle $\nu_e$~--~$\nu_s$ mixing consistent with the small mixing angle 
MSW solution to solar neutrino data.

There are however complications associated with a $U(2)$ family 
symmetry in supersymmetric theories.  It is believed that global 
symmetries do not arise in string theory and also these are thought to be 
violated by quantum gravity effects~\cite{qgravity_globsym}. On the other hand, 
with continuous gauge symmetries there are associated D -- term 
contributions to scalar masses which can lead to unacceptably large 
FCNCs~\cite{dterm}.  As a result, we should consider discrete family gauge 
symmetries.  Discrete gauge symmetries are not violated by quantum gravity 
effects~\cite{qgravity_discrgaugesym} and can arise in spontaneous 
breaking of continuous gauge symmetries or directly in compactifications of 
string theory. 

In this paper we present an $SO(10)$ SUSY GUT with $D_3 \times U(1)$ 
family gauge symmetry which does not suffer from the problems mentioned in 
the previous paragraph.  This model provides exactly the same operators 
generating Yukawa matrices as model~\cite{BRT}.
Thus it fits the low energy data in the charged lepton sector equally 
well and provides the same neutrino solution.  In addition, the field 
content of this model is simpler than~\cite{BRT} and can naturally provide 
an explanation for sequential family symmetry breaking by the vacuum 
expectation values [vevs] of ``flavon" fields.

The rest of the paper is organized as follows. In section 2 we briefly 
review possible discrete family symmetries, provide a motivation for $D_3 
\times U(1)$ as a family symmetry and discuss anomalies associated with 
gauging of this symmetry.  In section 3 we construct the 
$SO(10) \times D_3 \times U(1)$ invariant superspace potential which, 
after family symmetry breaking, generates the quark and lepton Yukawa 
matrices.  Our conclusions are in section 4.  For convenience, 
in Appendix A we summarize properties of the group $D_3$ and its 
representations, and calculate invariants used in section 3. 
In Appendix B we present a $D'_3$ version of the model and finally in 
Appendix C we briefly review the results of ~\cite{BRT} for charged fermion 
masses and mixing angles as well as for neutrino oscillations.

\section{Discrete Family Symmetry}

As mentioned in the introduction, we are interested in discrete family 
symmetries which posses two-dimensional irreducible representations. 
In order to be able to generate the same operators for fermion masses as 
in the case of $U(2)$, family symmetry~\cite{BRT} subgroups of $SO(3)$ or 
$SU(2)$ are suggested. 

Discrete subgroups of $SO(3)$ are classified~\cite{discrete_groups}
in terms of two infinite series: $Z_N$ (cyclic Abelian groups) and 
$D_N$ (non-Abelian dihedral groups); and three exceptional groups: 
$T$ (tetrahedral), $O$ (octohedral) and $I$ (icosahedral). Similarly, 
since $SO(3) \cong SU(2)/Z_2$, discrete subgroups of $SU(2)$ are classified 
in terms of double covers of the corresponding subgroups of $SO(3)$. We call 
these 
$Z'_N$, $D'_N$, $T'$, $O'$ and $I'$. Since $Z_N$ are abelian they posses 
only singlet irreducible representations. Irreducible representations of 
dihedral groups $D_N$ and $D'_N$ are all one and two dimensional. Three  
dimensional irreducible representations start to appear in the exceptional 
groups. 

In paper~\cite{BRT} the three generations of fermions transform as a doublet 
and singlet under $SU(2)$. To generate the effective mass operators for 
quarks and leptons in the light two generations, three ``flavon" fields 
$\phi^a$, $S^{ab}$ and $A^{ab}$ (doublet, symmetric triplet and 
anti-symmetric singlet under $SU(2)$) were introduced. The family symmetry 
is sequentially broken by {\em minimal} symmetry breaking vevs:
\begin{equation}
\langle \phi^a \rangle \; = \; \left( \begin{array}{c}
                               0\\
                               \phi \\
                               \end{array}  \right)  \; , \quad
\langle S^{ab} \rangle \; = \; \left( \begin{array}{cc}
                               0 & 0 \\
                               0 & S \\
                               \end{array}  \right)  \; , \quad
\langle A^{ab} \rangle \; = \; \left( \begin{array}{cc}
                               0 & A \\
                               - A & 0 \\
                               \end{array}  \right)  \; .
\label{eq:vevs}
\end{equation}

Thus, it looks like we need to consider a group which has at least one 
three dimensional irreducible representation to have a discrete analog of 
$S^{ab}$. In that case the tetrahedral group $T'$ would be the smallest 
group we could consider.~\footnote{In the process of writing this 
paper we became aware of the work~\cite{tetrahedral} which suggested 
the group $T'$ as a good starting point for models with ``$U(2)$-like" 
family symmetry.}  However, the coupling of a triplet to two doublets, 
which is necessary in~\cite{BRT}, can be easily mimicked by a coupling 
of three doublets 
in most of the dihedral groups. (In the case of $D_3$ see eqn. 
(\ref{eq:222}) in Appendix A and in the case of $D'_3$ eqn. (\ref{eq:2A2B2B}) 
in Appendix B.) Therefore a flavon field in the three-dimensional 
representation 
is not necessary when considering a dihedral family symmetry. Furthermore, 
it has not been possible to find a mechanism for generating non-zero vevs 
for $S^{22}$, while $\langle S^{11} \rangle = \langle S^{12} \rangle = 
0$~\cite{arash}.  On the other hand, if 
the most general family symmetry breaking vevs $\langle S^{11} \rangle = 
\k_1 S$,  $\langle S^{12} \rangle = \k_2 S$
are considered~\footnote{These new parameters have minor consequences in the
charged fermion sector ($\chi^2$ analysis requires them to be small), but 
provide new neutrino solutions.  For details see~\cite{BRT1}. } the 
predictivity of the theory is lost, since there are now as many parameters 
in the charged fermion sector as there are observables~\cite{BRT1}.

Therefore, dihedral groups are the most promising candidates for an
``$SU(2)$-like" family symmetry. They were previously used as family 
symmetries in refs.~\cite{dihedral}. If we now demand the minimal family 
symmetry group containing representations which can be 
used most economically, we are lead to the group $D_3$.

The group $D_3$ is the smallest non-Abelian group (it is isomorphic to $S_3$ - 
the symmetric permutation group).  Some basic properties of this group and its 
representations are summarized in Appendix A.  $D_3$ possesses three 
nonequivalent irreducible representations ${\bf 1_A}$, ${\bf 1_B}$ and 
${\bf 2_A}$ (${\bf 1_A}$ is a trivial representation; also denoted by
$1$). Thus this symmetry provides a natural interpretation of the three 
generations of fermions as a singlet and doublet ${\bf 1_B}$ + ${\bf 
2_A}$  under $D_3$. Differences between generations can then be understood 
as a consequence of assigning them to different representations of 
$D_3$.

Since we want the family symmetry to be gauged, it must be anomaly free. To 
show that there are no combined $D_3$ and/or $SO(10)$ anomalies we 
use the fact that both the $SO(3)$ and $SO(10)$ groups are anomaly free. 
Representations of $SO(3)$ decompose into irreducible representations 
of $D_3$ in the following way:
\begin{eqnarray}
 {\bf 1} & \rightarrow & {\bf 1_A} \; , \nonumber \\
 {\bf 3} & \rightarrow & {\bf 1_B} \; + \; {\bf 2_A} \; ,  \\    
 {\bf 5} & \rightarrow & {\bf 1_A} \; + \; {\bf 2_A} 
\; + \; {\bf 2_A} \; , \nonumber \\
 \vdots & & \nonumber 
\end{eqnarray}
Therefore, if the field content of the theory is such that fields with 
the same $SO(10)$ quantum numbers can be arranged into complete multiplets 
of $SO(3)$ then there are no $D_3$, $SO(10)$ or mixed anomalies. 

Because $D_3$ has only two nonequivalent nontrivial irreducible 
representations we also need (in order to maintain ``naturalness") an 
additional $U(1)$ symmetry to distinguish different fields with the 
same $D_3$ and $SO(10)$ charges.  This $U(1)$ symmetry is in general 
anomalous. An anomalous $U(1)$ gauge symmetry was previously used in 
models~\cite{anomU(1)}.  We shall assume that the $U(1)$ anomalies can be 
cancelled by the Green -- Schwarz mechanism~\cite{Green-Schwarz}. 

Before we continue, it is important to discuss the consequences of the 
symmetry group $D_3$ with regards to flavor violation~\cite{flavorviolation}.
It has been shown that an SU(2) family symmetry can effectively suppress
flavor violating processes among the first two 
families~\cite{u2symmetry,so10u2}.
This follows from the fact that to zeroth order in family symmetry breaking,
the soft SUSY breaking mass term for squarks and sleptons in the first two
families is an SU(2) invariant and thus proportional to the identity matrix.
Then family symmetry breaking corrections to squark and slepton masses are
at most of order the family mixing for quarks and leptons.   In appendix A, 
we show that the same argument also applies for $D_3$.  Thus $D_3$ will also
suppress flavor violations.

\section{An $SO(10) \times D_3 \times U(1)$ Model}

In this section we present an $SO(10)$ SUSY GUT with $D_3 \times U(1)$ 
family gauge symmetry. In $SO(10)$ all fermions in one generation are 
contained in the 16 dimensional irreducible representation and, in the 
simplest version, one pair of Higgs doublets is contained in the 10 
dimensional irreducible representation. The minimal Yukawa coupling of 
the third generation of fermions to the Higgs fields is given by $\lambda 
16_3 10 16_3$ from which we obtain the symmetry relation $\lambda_t = 
\lambda_b = \lambda_{\tau} = \lambda_{\nu_{\tau}} = \lambda$ at the GUT 
scale. While this Yukawa unification is known to work quite well for the 
third generation it fails for the two light generations. 
Thus a family symmetry is necessary to forbid the tree level Yukawa 
coupling of the first and second generations to the Higgs fields. 
Breaking of this symmetry will provide the necessary hierarchy of fermion 
masses.

\subsection{The charged fermion sector}

As discussed in section 2 the first two generations of fermions are 
contained in $16_a$, $ a = 1,2$ which is a doublet under $D_3$ with 
charge $1$ under $U(1)$ [ or $16_a = ({\bf 2_A}, 1)$ ]. 
The third generation $16_3$ transforms as $({\bf 1_B}, 3)$ and a 
$10$ of Higgs transforms as $(1, -6)$. 
Using the results of Appendix A we see that the coupling $\lambda 16_3 10 
16_3$ is invariant under $D_3 \times U(1)$ while $\lambda 16_a 10    
16_a$ and $\lambda 16_a 10 16_3$ are not.

To generate the Yukawa couplings for the first two generations we 
introduce three ``flavon" superfields: 
\begin{equation}
\p = ({\bf 2_A}, -2) \; , \nonumber
\quad \ptil = ({\bf 2_A}, -4) \; , \nonumber
\quad A = ({\bf 1_B}, 4) \; , \nonumber
\end{equation}
which are $SO(10)$ singlets, and a pair of Froggatt-Nielsen 
states~\cite{FN} ($\overline {16}$ and $16$ under $SO(10)$): 
\begin{equation}
\bar \chi_a = ({\bf 2_A}, -5) \; , \nonumber
 \quad \chi_a = ({\bf 2_A}, 5) 
\; . 
\nonumber
\end{equation}

The superspace potential for the charged fermion sector of this model is 
given by: 
\begin{eqnarray}  
W \supset \; \; &16_3 \; 10\; 16_3 & + \; \;  16_a \; 10 \; \chi_a 
\nonumber \\
 + & \bar \chi_a  \; ( \; M \; \chi_a  & 
+ \; \; \frac{1}{M_0} 
\; 45 \; \ptil 16_3 \; \; + \; \; \frac{1}{M_0} \; 45 \; \p 16_a \; \;
+ \; \;  A \; 16_a \; ) \; ,
\label{eq:Wchf}
\end{eqnarray}
where $45 = (1, 6)$ is an $SO(10)$ adjoint field~\footnote{Note that we 
usually call fields by their SO(10) quantum numbers. The adjoint 
representation of $SO(10)$ is 45 dimensional.} which is assumed to obtain 
a vev in the B -- L direction; and $M = (1, 0)$ is a linear combination of an 
$SO(10)$ singlet and adjoint. Its vev $M_0 ( 1 + \a X + \b Y)$ gives mass 
to Froggatt-Nielsen states. Here $X$ and $Y$ are elements of the Lie 
algebra of $SO(10)$ with $X$ in the direction of the $U(1)$ which 
commutes with $SU(5)$ and $Y$ the standard weak hypercharge; and $ \a $ , 
$ \b $ are arbitrary constants which are fit to the data. Furthermore, each 
term in $W$ has an arbitrary coupling constant which is omitted for 
notational simplicity.~\footnote{To forbid all higher dimensional 
operators we also assume a $U(1)_R$ symmetry under which $45$ has zero 
charge and all other fields have charge $1$. Neither the $U(1)$ nor $U(1)_R$ 
symmetry is in any sense unique. We can equally well assume just one 
symmetry without imposing R-symmetry or products of several $U(1)$s or 
their discrete subgroups $Z_N$. By specifying $U(1)$ charges we show that 
the model is ``natural," i.e. there exist a $U(1)$ which allows the required 
operators in the superpotential and at the same time forbids all possibly 
dangerous operators to any order. If we do not impose the $U(1)_R$ symmetry 
the model is however still natural.  The charges under any single 
$\tilde U(1)$ 
which constrains the model are however relatively high; a reflection of the 
fact that this symmetry has to forbid all dangerous higher dimensional 
operators. An example of such a $\tilde U(1)$ is (including fields which 
occur later in sections 3.2 and 3.3): $16_a = -4$, $16_3 = -3$, $10 = 6$,
$\bar \chi_a = -16$, $\chi_a = -2$, $\ptil = 7$, $\p = 8$, $A = 20$, $45 
= 12$, $M = 18$, $\psi_a = -15$, $S = -15$, $S_{\phi} = 15$,
$N_a = -4$, $N_3 = -3$, and $\overline {16} = 6$.} 

\begin{center}   
\begin{figure}
\centerline{ \psfig{file=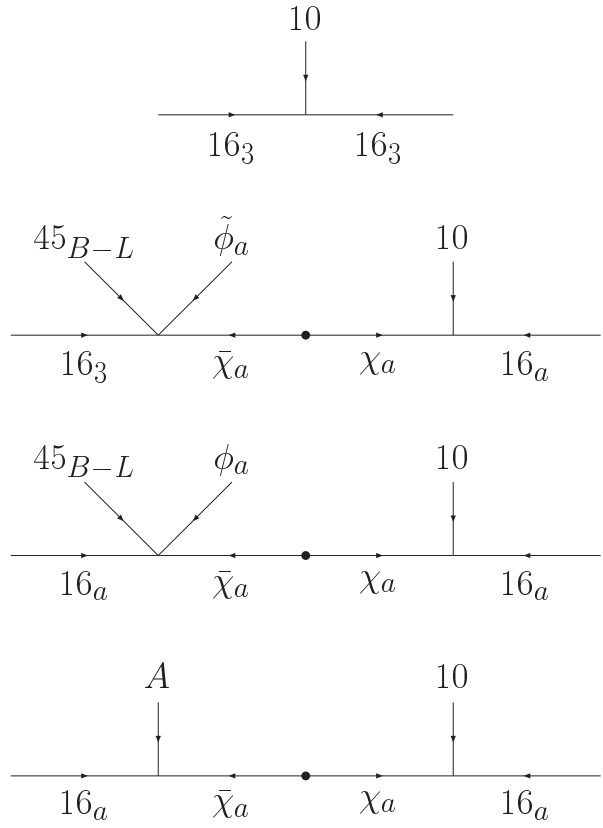,width=12cm,rheight=13cm}}
\caption{\it Diagrams generating the Yukawa matrices.}
\label{figure:twodoublets}
\end{figure}
\end{center}

The largest scale of the theory is assumed to be the mass of the 
Froggatt-Nielsen states. In the effective theory below $M_0$, these states 
are integrated out giving the effective mass operators 
in Figure~\ref{figure:twodoublets}.

When ``flavon" doublets obtain vevs $\langle \ptil \rangle = \tilde\phi \; 
\d_{a 1}$ and $\langle \p \rangle = \phi \; \d_{a 2}$ the family symmetry 
$D_3 \times U(1)$ is broken to a diagonal $Z_6$ symmetry and the Yukawa 
couplings $16_3 ... 16_2$ and $16_2 ... 16_2$ are generated. Finally, 
the vev of the $A$ field breaks the family 
symmetry completely and generates the Yukawa coupling 
$16_1 ... 16_2$. These results are summarized in the form of the Yukawa 
matrices for up quarks, down quarks, charged leptons and the Dirac 
neutrino Yukawa matrix below.~\footnote{The ratios of vevs which enter 
the Yukawa matrices are given by dimensionless parameters: 
$r \epsilon \sim \tilde\phi \; \langle 45 \rangle  
\, / \, M_0^2$,
$\epsilon \sim \phi \; \langle 45 \rangle \, / \, M_0^2 $,
$\epsilon' \sim A \, / M_0 $. Parameters $\sigma$ and $\rho$ are 
functions of $\alpha$ and $\beta$ which were defined after equation 
(\ref{eq:Wchf}). For more details see~\cite{so10u2}.  }
\begin{eqnarray}
Y_u =&  \left(\begin{array}{ccc}  0 & \epsilon' \rho & 0 \\
                          - \epsilon' \rho &  \epsilon \rho & r \epsilon
T_{\bar   
u}     \\
                      0  & r \epsilon T_Q& 1 \end{array} \right) \; 
\lambda &
\nonumber \\
Y_d =&  \left(\begin{array}{ccc}  0 & \epsilon'  & 0 \\
                          - \epsilon'  &  \epsilon  &  r \sigma \epsilon
T_{\bar
d}\\
                      0  &  r \epsilon T_Q & 1 \end{array} \right) \; \xi
& \label{eq:yukawa}
  \\
Y_e =&  \left(\begin{array}{ccc}  0 & - \epsilon'  & 0 \\
                           \epsilon'  &  3 \epsilon  &  r \epsilon T_{\bar
e} \\
                      0  &  r \sigma \epsilon T_L & 1 \end{array} \right)
\; \xi &
 \nonumber \\
Y_{\nu} =&  \left(\begin{array}{ccc}  0 & - \omega \epsilon'  & 0 \\
                  \omega \epsilon'  &  3 \omega \epsilon  & {1 \over 2}
\omega r
\epsilon T_{\bar \nu}
\\
                      0  &  r \sigma \epsilon T_L& 1 \end{array} \right) \;
\lambda &
 \nonumber
\end{eqnarray}
with  \begin{eqnarray} \omega = {2 \, \sigma \over 2 \, \sigma - 1}
\label{eq:omega} \end{eqnarray} and
\begin{eqnarray} T_f  = & (\rm Baryon\# - Lepton \#) &
\label{eq:Tf} \\
\rm for & f = \{Q,\bar u,\bar d, L,\bar e, \bar \nu\}.& \nonumber
\end{eqnarray}

In our notation, fermion doublets are on the left and singlets are on the
right.  Note, we have assumed that the Higgs doublets of the minimal 
supersymmetric standard model[MSSM] are contained in the $10$ such that
$\lambda \;10 \supset \lambda \; H_u \; + \; \xi \; H_d$. We could then
consider two important limits ---  case (1)  $\lambda = \xi$ (no Higgs 
mixing) with large $\tan\beta$, and case (2)  $\lambda \gg \xi$ or small 
$\tan\beta$.  In the first case the Yukawa matrices are given by specifying 
six real parameters $\lambda , \e , \e' , \r , \s , r $ and three phases 
$\Phi_{\e}, \Phi_{\s}, \Phi_{\r}$, which cannot be rotated away. 
These nine parameters are then fit to the thirteen observable 
charged fermion masses and mixing angles.  In the second case we would have 
one more arbitrary parameter.

We have obtained the Yukawa matrices parameterized in the same way as in 
the paper~\cite{BRT}. Therefore, all the results from~\cite{BRT} apply 
also in our case.  For completeness, the results for charged fermion masses 
and mixing angles are summarized in the Appendix C.

\subsection{The Superpotential for ``flavon" doublets}

To generate the Yukawa matrices (\ref{eq:yukawa}) with zeros in the 1 -- 1, 1 
-- 3 and 3 -- 1 elements it is necessary to have $\langle \tilde\phi_2 
\rangle \, = \, \langle \phi_1 \rangle \, = 0$. This may look like a very 
special assumption. However, we argue that with a $D_3$ symmetry such an 
arrangement of vevs for ``flavon" doublets is naturally obtained.

Consider the following superpotential for ``flavon" doublets:
\begin{equation}
W \supset \; \; \psi_a \; \p \; \ptil \; \; + \; \; S \; 
\left( \; \p \; \ptil \; \; - \; \; M^2_{\phi} \; \right) \; ,
\label{eq:Wflavon}
\end{equation}
where $\psi_a = ({\bf 2_A}, 6)$ and $S = (1, 6)$ are singlets under 
$SO(10)$. $M^2_{\phi}$ is a scale at which the ``flavon" doublets obtain 
vevs. It is effectively $(1, -6)$. The origin of $M^2_{\phi}$ is not 
important. It can result from one or two fields with effective $U(1)$ 
charge $-6$ obtaining a vev. For example, if $M^2_{\phi} = \lambda_{\phi} 
\langle S_{\phi} \rangle $, where $\lambda_{\phi}$ is a dimensionful 
constant, it can be checked 
that $S_{\phi} = (1, -6)$~\footnote{ $S_{\phi}$ has charge $2$ 
under $U(1)_R$ symmetry} does not couple anywhere else; neither in the 
charged lepton sector nor the neutrino sector 
(see next section).~\footnote{When $S_{\phi}$ obtains 
a vev, the $U(1)$ symmetry is broken down to $Z_{6}$. As we saw in the
previous section the vevs of $\p$ and $\ptil$ leave an unbroken $Z_6$ symmetry.
Therefore, to be precise, with this mechanism for generating appropriate 
vevs of ``flavon" doublets the flavor symmetry breaking scenario from the 
previous section is slightly changed to $D_3 \times U(1) \rightarrow D_3 
\times Z_{6} \rightarrow Z_6 \rightarrow$ nothing.} 

The superpotential (\ref{eq:Wflavon}) has two isolated supersymmetric 
vacua related by $\p \leftrightarrow \ptil$:
\begin{equation}
\psi_a \; = \; S \; = \; 0 \; , \quad \p \; = \; \left( \begin{array}{c}
                                                       0 \\
                                                       \phi\\
                                                       \end{array}  \right)
\; , \quad \ptil \; = \; \left( \begin{array}{c}
                               \tilde\phi \\
                               0 \\
                               \end{array}  \right)  \; ,
\quad \phi \tilde\phi \; = \; M^2_{\phi} \; .
\end{equation}
Since $\psi_a$ and $S$ have zero vevs they do not contribute in the 
charged lepton and the neutrino sectors.

Thus from the simple superpotential (\ref{eq:Wflavon}) we have obtained 
the solution for vevs of $\p$ and $\ptil$ needed to generate 
the Yukawa matrices (\ref{eq:yukawa}).

\subsection{The Neutrino sector}

The parameters in the  Dirac Yukawa matrix for neutrinos 
(\ref{eq:yukawa}) mixing $\nu - \bar \nu$ are now fixed.  
Of course, neutrino masses are much too large and we need to invoke the 
GRSY~\cite{grsy} see-saw mechanism.

We can introduce SO(10) singlet fields $N$ and obtain effective mass 
terms $\bar \nu - N$ and $N - N$.  Adding $N_a = ({\bf 2_A}, 1)$ 
and $N_3 = ({\bf 1_B}, 3)$ (with the same $U(1)$ charges as $16_a$ 
and $16_3$) together with $\overline {16} = (1, -6)$ 
(the same $U(1)$ charge as $10$)~\footnote{$\overline {16}$ is 
assumed to get a vev in the ``right-handed" neutrino direction. 
This vev is also needed to break the rank of $SO(10)$.} we directly 
obtain the terms $\bar \nu - N$.  The corresponding diagrams can be 
obtained from Figure~\ref{figure:twodoublets} by substituting 
$10 \rightarrow \overline{16}$, $16_a \rightarrow N_a$, $16_3 \rightarrow 
N_3$ on the right hand side of the diagrams. $N - N$ mass terms are generated 
from operators describing interactions of $N_a$ and $N_3$ with flavon fields.
Thus these new fields contribute to the superspace potential below.
\begin{equation}
W \supset \overline{16} \; \left( \; N_a \; \chi_a \;\; + \;\; N_3 \; 
16_3 \right) \;\;
 + \; \; N_a \; N_a \; \p \;\; + \;\; N_a \; N_3 \; \ptil \; .
\end{equation}

Finally in order to allow for the possibility of a light sterile neutrino 
we introduce a $D_3$ nontrivial singlet $\bar N_3$ (a singlet 
under $SO(10)$) which enters the superspace potential as follows.
\begin{equation}
W \; \supset \;\mu_3 \;  N_3 \; \bar N_3  \label{eq:mu'}
\end{equation}
The dimensionful parameter $\mu_3$ is assumed to be of order the
weak scale.  The notation is suggestive of the similarity between this 
term and the $\mu$ term in the Higgs sector. In both cases, we are 
adding supersymmetric mass terms and in both cases, we need some mechanism 
to keep these dimensionful parameters small compared to the Planck scale. 
This may be accomplished by symmetries, see for example 
ref.~\cite{giudicemasiero}.

We define the vector $\tilde \mu = ( 0, 0, \mu_3 )^T $
which can be generalized to a matrix in the case of more than one sterile 
neutrino.

The case with three neutrinos ($\mu_3 \equiv 0$) cannot simultaneously fit 
both solar and atmospheric neutrino data, for details see~\cite{BRT}. In 
this paper we consider the case of four neutrinos (with one sterile neutrino).

The generalized neutrino mass matrix is then given by:~\footnote{This
is similar to the double see-saw mechanism suggested by Mohapatra and    
Valle~\cite{mohapatra}.}

\begin{eqnarray}
& ( \begin{array}{cccc}\; \nu & \;\; \bar N_3 & \;\; \bar \nu & \;\;  N
\end{array})  &
\nonumber\\  &  \left( \begin{array}{cccc}  0 & 0 & m & 0  \\
                     0 & 0 & 0 & \tilde \mu^T \\
                     m^T & 0 & 0 & V \\
                     0 & \tilde \mu & V^T & M_N  \end{array} \right) &
\end{eqnarray}
where  \begin{eqnarray} m = & Y_{\nu}\; \langle H_u^0 \rangle
&= \; Y_{\nu}\; {v \over\sqrt{2}}\; \sin\beta \end{eqnarray} and
\begin{equation}
V \; = \; \left( \begin{array}{ccc}  0 & \epsilon' V_{16} & 0 \\
                                - \epsilon' V_{16} & 3 \epsilon V_{16} & 
0 \\
                                 0  & r \, \epsilon \, (1 - \sigma) \,
T_{\bar \nu} V_{16}  &
V'_{16}
\end{array}\right) \; , \quad 
  M_N = \left( \begin{array}{ccc}  0 & 0 & 0 \\
                                0 & \phi & \tilde \phi \\
                                 0  & \tilde \phi  &  0 
               \end{array}\right) \; . 
\end{equation}
$V_{16},\; V'_{16}$ are proportional to the vev of $\overline{16}$
(with different implicit Yukawa couplings) and $\phi, \; 
\tilde \phi$ are up to couplings the vevs of $\phi_2, \; \tilde \phi_1$, 
respectively.

Since both $V$ and $M_N$ are of order the GUT scale, the states $\bar \nu,
\; N$ may be integrated out of the effective low energy theory.  In this 
case, the effective neutrino mass matrix is given (at $M_G$) 
by~\footnote{In fact, 
at the GUT scale $M_G$ we define an effective dimension 5 supersymmetric 
neutrino mass operator where the Higgs vev is replaced by the Higgs 
doublet  H$_u$ coupled to the entire lepton doublet.  This effective 
operator is then renormalized using one-loop renormalization group 
equations to $M_Z$.  It is only then that $H_u$ is replaced by its vev.} 
(the matrix is written in the ($\nu, \ \bar N_3$) {\em flavor} basis 
where  charged lepton masses are diagonal).
\begin{equation}
 m^{eff}_{\nu} \; = \;  \tilde U_e^{\dag} \; \left( \begin{array}{cc}
m\;(V^T)^{-1}\;M_N\; V^{-1}\; m^T &  - m \;(V^T)^{-1}\; \tilde \mu\\
                    - {\tilde \mu}^T \; V^{-1}\; m^T & 0  \end{array}
\right) \; \tilde U_e^{\ast}
\end{equation}
with
\begin{equation}
\tilde U_e \; = \; \left(\begin{array}{cc} U_e & 0 \\
                                  0 & 1 \end{array}\right) \; , \quad
e_0 \; = \; e \; U_e^{\dag} \; , \quad  \nu_0 \; = \; \nu \; U_e^{\dag} \; .
\end{equation}
$U_e$ is the $3\times3$ unitary matrix for left-handed leptons needed to
diagonalize $Y_e$ (eqn. ~\ref{eq:yukawa}) and $e_0,\; \nu_0 \; (e, \; 
\nu)$  represent the three families of left-handed leptons in the 
weak- (mass-) eigenstate basis for charged leptons.

The neutrino mass matrix is diagonalized by a unitary matrix $U =
U_{\alpha\, i}$;
\begin{equation}
m^{diag}_{\nu} \; = \; U^{\dag} \; m_{\nu}^{eff} \; U^{\ast} 
\end{equation}
where $\alpha= \{\nu_e ,\; \nu_\mu ,\; \nu_\tau ,\; \nu_{s} \}$
is the flavor index and
$i = \{ 1, \cdots, 4\}$ is the neutrino mass eigenstate index.
$U_{\alpha\, i}$  is observable in neutrino oscillation experiments.  
In particular,  the probability for the flavor state $\nu_\alpha$ with 
energy $E$ to oscillate into $\nu_\beta$ after traveling a distance $L$ 
is given by
\begin{equation}
P(\nu_\alpha \rightarrow \nu_\beta) \;  = \; \delta_{\alpha \beta}
\; \; - \; \; 4\sum_{k\, <\, j} U_{\alpha \, k} \ U^*_{\beta \, k} \
U^*_{\alpha \, j} \ U_{\beta \, j} \ \sin^2\Delta_{j\, k} \; , 
\end{equation}
where $\Delta_{j\,k} =  {\delta m^2_{j k} \ L \over 4 E}$ and
$\delta m^2_{j k} = m^2_j - m^2_k$.

The results for this four neutrino model (taken from ref.~\cite{BRT}) are given
in Appendix C.

\subsection{Anomalies}

As mentioned in section 2, we restrict discussion of anomalies to
those involving $D_3$ and $SO(10)$ only. The only fields in the model 
with nontrivial charge under both groups are: doublets $16_a$, $\chi_a$,
$\bar \chi_a$ and ${\bf 1_B}$ singlet $16_3$. The simplest way to avoid 
anomalies is to arrange these fields into ${\bf 2_A} + {\bf 1_B}$ multiplets
of $D_3$ with the same $SO(10)$ quantum number. To make this possible we have to
introduce another pair of Froggatt-Nielsen fields $\chi$ and $\bar \chi$
which are ${\bf 1_B}$ singlets under $D_3$. It is easy to check that
these new fields do not contribute to the discussion in this section.

There are many ways to arrange the $SO(10)$ singlets with non-trivial 
$D_3$ quantum numbers
into complete multiplets of $SO(3)$. In particular, it is always possible to 
add new doublets or ${\bf 1_B}$ singlets under $D_3$ which do not 
contribute to
fermion masses and mixing angles.

In Appendix B we present a $D'_3$ version of the same model. The main
advantage of $D'_3$ is that the ${\bf 2}$ of $SU(2)$ decomposes into the ${\bf 
2_B}$ representation of $D'_3$. Thus, if all doublets with nontrivial
$SO(10)$ quantum numbers transform as ${\bf 2_B}$ under $D'_3$ the 
anomaly cancellation conditions are automatically satisfied.

\section{Conclusions}
In this paper we have presented an SO(10) SUSY GUT with the minimal discrete 
non-abelian gauge family symmetry, $D_3 \times U(1)$.\footnote{As mentioned 
in section 3.1, the U(1) factor can even be replaced by a discrete $Z_N$ 
symmetry.}   With minimal family symmetry breaking vevs, which may be 
obtained naturally in this theory, we 
obtain a ``predictive" model for quark and lepton masses 
(including neutrinos) which will be tested in future experiments.   
In the charged fermion sector the model reproduces the good results obtained 
previously in an SO(10)$\times$U(2)$\times$U(1) model discussed 
in ref.~\cite{BRT}.  The $D_3$ symmetry is sufficient to suppress large 
flavor violating interactions in the charged fermion sector.  In the neutrino 
sector we 
also reproduce the results of ref.~\cite{BRT}, in particular we are able 
to fit atmospheric neutrino data with maximal $\nu_\mu \rightarrow \nu_\tau$ 
oscillations and solar neutrino data with SMA MSW $\nu_e \rightarrow \nu_s$ 
oscillations.  The model is however unable to fit LSND data.

\noindent
{\bf Acknowledgements} This work was supported in part by DOE/ER/01545-772 
and partially by a Fermilab Frontier Fellowship.   We would like to thank the 
Fermilab Theory Group for their kind hospitality during our stay there.
We would also like to thank A. Mafi for discussions and T. Bla\v zek and 
K. Tobe for the use of the work in preparation.  

\section*{Appendix A. The group $D_3$ and its representations}

\begin{center}
\begin{figure}[h]

\vspace*{-4cm}
\centerline{ \psfig{file=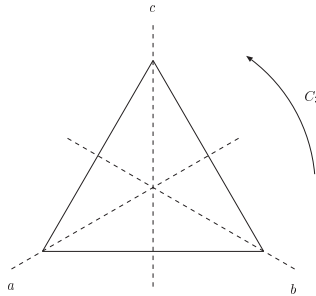,width=12cm,rheight=12cm}}

\vspace*{-5cm}

\caption{\it Symmetry axes of an equilateral triangle.}
\label{figure:D3triangle}

\end{figure}
\end{center}

All possible rotations in three dimensions which leave an equilateral 
triangle invariant form the group $D_3$ (see Figure~\ref{figure:D3triangle}). 
This group contains six elements 
in three classes:~\footnote{An element $b$ of the group $G$ is said to be 
{\it conjugate} to the element $a$ if there is an element $u$ in $G$ such 
that $u a u^{-1} = b$. A group can be separated into {\it classes} of 
elements which are conjugate to one another.}
\begin{equation}
E; \; \;  C_3, \, C_3^2 ; \; \; C_a,\, C_b, \, C_c, 
\nonumber
\end{equation}
\noindent
where $E$ is the identity element, $C_3$ is the rotation through $2\pi 
/3$ about the axis perpendicular to the paper and going through the 
center of the triangle, $C_3^2$ is $C_3$ applied twice, $C_a$ is the 
rotation through $\pi$ about the axis $a$, and similarly $C_b$ and $C_c$. 
Note that $C_b$ is the same as $C_a C_3$ 
and $C_c$ is the same as $C_a C_3^2$.
\noindent


The number of classes in a finite group is equal to the number of 
nonequivalent irreducible representations of the group. One of the most 
interesting results of the theory of finite groups is the relation 
between the number of elements $g$ of a group and dimensions $n_{\nu}$ of 
its nonequivalent irreducible representations $\nu$,\\
$$ \sum_{\nu} n_{\nu}^2 \, = \, g. $$

\noindent
Thus we find that the group $D_3$ has two nonequivalent one dimensional 
representations ${\bf 1_A}$, ${\bf 1_B}$ and one two dimensional 
representation ${\bf 2_A}$. Each representation is described by the set of 
characters~\footnote{The character of an element $a$ of the group $G$ in 
a given representation $D$ is the trace $\sum_i D_{ii}(a)$. Therefore 
elements in the same class (conjugate elements) have the same character.}
$\chi_1,  ...,  \chi_{\nu}$, where $\nu$ is the number of classes in 
the group. The character table for the group $D_3$ is given in 
Table~\ref{table:D3}.

\begin{table}[h]
$$
\begin{array}{|c|c|c|c|}
\hline 
\quad D_3  \quad & \quad E \quad  & \quad C_3 \quad & 
\quad C_a \quad \\ \hline
\hline
{\bf 1_A}   &  1  &  1    &  1    \\
{\bf 1_B}   &  1  &  1    &  -1   \\
{\bf 2_A}     &  2  &  -1   &  0     \\
\hline
\end{array}
$$

\caption{\it The character table for the group $D_3$.}
\label{table:D3}
\end{table}

From the character table it is possible to find the decomposition of the 
product of any two representations:
\begin{equation}
{\bf 1_A} \otimes {\bf 1_A} = {\bf 1_A}, \quad {\bf 1_A} \otimes {\bf 1_B} = 
{\bf 1_B}, \quad {\bf 1_B} \otimes {\bf 1_B} = {\bf 1_A},
\label{eq:1x1}
\end{equation}

\begin{equation}
{\bf 1_A} \otimes {\bf 2_A} = {\bf 2_A}, \quad {\bf 1_B} \otimes 
{\bf 2_A} = {\bf 2_A}, 
\label{eq:1x2}
\end{equation}

\begin{equation}
{\bf 2_A}  \otimes {\bf 2_A} = {\bf 1_A} \oplus {\bf 1_B} \oplus 
{\bf 2_A}.
\label{eq:2x2}
\end{equation}

To construct an explicit model obeying $D_3$ symmetry we need to specify 
the representation and determine invariant tensors.
One dimensional representations coincide with the characters and the two 
dimensional representation can be chosen to be:
\begin{equation}
D(E) = \left( \begin{array}{cc}
1 & 0 \\ 0 & 1 \end{array}  \right) \; , \quad 
D(C_3) = \left( \begin{array}{cc}
\e & 0 \\ 0 & \e^{-1} \end{array}  \right) \; , \quad
D(C_a) = \left( \begin{array}{cc}
0 & 1 \\ 1 & 0 \end{array}  \right) \; ,
\end{equation}

\noindent
where $\e = e^{2\pi i/3}$. 

Now it is straightforward to find the two singlets and the doublet in the 
decomposition of a product of two doublets (\ref{eq:2x2}).
Writing $\psi = \{ x, y \}$ and $\psi ' = \{ x', y' \}$, we find:

\begin{equation}
\psi \otimes \psi ' \; |_{1_A} = x y' + y x' \; ,      \label{eq: 22A}
\end{equation}

\begin{equation}
\psi \otimes \psi ' \; |_{1_B} = x y ' - y x' \; ,     \label{eq: 22B}
\end{equation}

\begin{equation}
\psi \otimes \psi ' \; |_2 = \left( \begin{array}{c}
                                  y y' \\
                                  x x' \\
                                  \end{array}  \right) .
\end{equation}

The decomposition (\ref{eq:2x2}) also reveals that the product of three 
doublets contains an invariant. Taking $\psi '' = \{ x '' , y '' \}$, 
this invariant is:

\begin{equation}
\psi \otimes \psi ' \otimes \psi '' \; |_{1_A} = x x' x'' + y y' y'' \; .
\label{eq:222}
\end{equation}

Finally, we want to show that given a doublet $\psi_a$ in $D_3$, there 
is a unique invariant norm given by  $\psi_a^* \psi_a \equiv \psi_1^* 
\psi_1 + \psi_2^* \psi_2$.   Clearly, this norm is $D_3$ invariant since 
under a $D_3$ transformation  $\psi_a^\prime = C_{a b} \psi_b$ 
with $C \subset D_3$ and  $C^\dagger C = 1$.   That this is unique follows 
from the fact that in the product of two doublets there is a unique invariant 
given in eqn. (\ref{eq: 22A}).
In addition, defining a new doublet by $\chi_a = g_{a b} \psi_b^*$ satisfying
$\chi_a^\prime = C_{a b} \chi_b = (\psi_b^*)^\prime g_{b a}^T = 
\psi_c^* C_{c b}^\dagger g_{b a}^T$ requires for consistency 
$ g = C g C^T$.  The unique solution to this consistency condition 
is  $g = \left( \begin{array}{cc} 0 & 1 \\ 1 & 0
\end{array} \right)$.   
Then we have $\chi \otimes \psi  \; |_{1_A} \equiv \psi_a^* \psi_a$.

\section*{Appendix B. $D'_3$ version of the model}

The double group $D'_3$ contains 12 elements in 6 classes. 
In addition to ${\bf 1_A}$, ${\bf 1_B}$, and ${\bf 2_A}$ representations 
which are already presented in $D_3$ it also has double-valued 
representations ${\bf 1_C}$, ${\bf1_{\bar C}}$ and ${\bf 2_B}$. The 
character table of the double-valued representations is given in 
Table~\ref{table:D'3}. 
\begin{table}[h]
$$
\begin{array}{|c|c|c|c|c|c|c|}
\hline
\quad D_3  \quad & \quad E \quad  & \quad R \quad  &
\quad C_3 \quad & \quad C_3 R \quad &
\quad C_a \quad  & \quad C_a R \quad   \\
\hline
\hline
{\bf 1_C}         &  1  &  -1    &  -1 &  1  &  i   &  -i    \\
{\bf 1_{\bar C}}  &  1  &  -1    &  -1 &  1  &  -i  &  i     \\
{\bf 2_B}         &  2  &  -2    &   1 &  -1 &  0   &  0     \\
\hline
\end{array}
$$
\caption{\it The character table for double-valued representations 
of the group $D'_3$.} 
\label{table:D'3}
\end{table}

\noindent
Multiplication rules are given in Table~\ref{table:mult.rules} and 
equations (\ref{eq:2x2}), (\ref{eq:2x2'}) and (\ref{eq:2'x2'}).
\begin{table}[h]
$$
\begin{array}{|c|c|c|c|c|c|c|}
\hline
\quad D'_3  \quad & \quad {\bf 1_A} \quad  & \quad {\bf 1_B} \quad &
\quad {\bf 1_C} \quad & \quad {\bf 1_{\bar C}} \quad  & \quad {\bf 2_A} 
\quad & {\bf 2_B} \quad \\
\hline
\hline
{\bf 1_A}   & {\bf 1_A}  & {\bf 1_B} &  {\bf 1_C} & {\bf 1_{\bar C}} &
{\bf 2_A} & {\bf 2_B} \\
{\bf 1_B} & & {\bf 1_A}  & {\bf 1_{\bar C}} &  {\bf 1_C} & 
{\bf 2_A} & {\bf 2_B} \\
{\bf 1_C} & & & {\bf 1_B} & {\bf 1_A}  & {\bf 2_B}  & {\bf 2_A} \\
{\bf 1_{\bar C}} & & & & {\bf 1_B} & {\bf 2_B}  & {\bf 2_A} \\
\hline
\end{array}
$$
\caption{\it Multiplication rules for the group $D'_3$.} 
\label{table:mult.rules}
\end{table}

\begin{equation}
{\bf 2_A} \otimes {\bf 2_B} = {\bf 1_C} \oplus {\bf 1_{\bar C}}
\oplus {\bf 2_B} \; , 
\label{eq:2x2'}                
\end{equation}                    
\begin{equation}
{\bf 2_B} \otimes {\bf 2_B} = {\bf 1_A} \oplus  {\bf 1_B} \oplus {\bf 
2_A} \; .
\label{eq:2'x2'}
\end{equation}

\noindent
The double-valued two dimensional representation can be chosen to be:
\begin{equation}
D(E) = \left( \begin{array}{cc}
1 & 0 \\ 0 & 1 \end{array}  \right) \; , \quad
D(C_3) = \left( \begin{array}{cc}
\e^{1/2} & 0 \\ 0 & \e^{-1/2} \end{array}  \right) \; , \quad
D(C_a) = \left( \begin{array}{cc}
0 & 1 \\ -1 & 0 \end{array}  \right) \; ,
\end{equation}
and $D(R) = - D(E)$. As before, $\e = e^{2\pi i/3}$.

Now it is straightforward to find new invariants. Taking the ${\bf 2_A}$
doublet 
$\psi = \{ x, y \}$ and ${\bf 2_B}$ doublets $\varphi = \{ a, b \}$, 
$\varphi' = \{ a', b' \}$ we find:
\begin{equation}
\varphi \otimes \varphi' \; |_{1_A} \; = \; a b' - b a' \; ,      
\label{eq: 2A2B}
\end{equation}
\begin{equation}
\psi \otimes \varphi \otimes \varphi'\; |_{1_A}  \; = \;
x b b' + y a a' \; .
\label{eq:2A2B2B} 
\end{equation}

With these results it is straightforward to check that the fermion masses 
and mixing angles we obtained in section 3 can be also obtained if we 
assume a $D'_3 \times U(1)$ family symmetry. In this case all doublets 
charged nontrivially under $SO(10)$ are in the ${\bf 2_B}$ of $D'_3$, while 
singlets transform trivially under $D'_3$. 
``Flavon" fields are in representations: $ \p = {\bf 2_A}$, $\ptil = {\bf 
2_B}$ and $A = {\bf 1_A}$. ``Flavon" doublets are expected to obtain vevs
$\langle \p \rangle = \phi \; \d_{a 1}$ and $\langle \ptil \rangle = 
\tilde\phi \; \d_{a 1}$.
 
In the neutrino sector the doublets transform in the
${\bf 2_B}$ and the singlets transform trivially under $D'_3$.
Finally, the fields entering the superpotential for the ``flavon" doublets 
transform in the following way: $\psi_a = {\bf 2_B}$, $S = {\bf 1_C}$,
and  $S_{\phi} = {\bf 1_{\bar C}}$.

The advantage of $D'_3$ (and $D'_N$s in general) is that the ${\bf 2_B}$ 
representation of $D'_3$ appears alone in the decomposition of a ${\bf 2}$ 
of $SU(2)$.  Representations of $SU(2)$ decompose into irreducible 
representations
of $D'_3$ in the following way:
\begin{eqnarray}
 {\bf 2} & \rightarrow & {\bf 2_B} \; , \nonumber \\
{\bf 3} & \rightarrow & {\bf 1_B} \; + \; {\bf 2_A} \\
 {\bf 4} & \rightarrow & {\bf 1_C} \; + \; {\bf 1_{\bar C}} \; + 
\; {\bf 2_B} \; ,  \\
 \vdots & & \nonumber
\end{eqnarray}
Because all doublets with nontrivial $SO(10)$ quantum numbers transform as 
${\bf 2_B}$ under $D'_3$ and all singlets with nontrivial $SO(10)$ 
quantum numbers are trivial singlets under $D'_3$ the anomaly cancellation 
conditions are automatically satisfied.   For the $SO(10)$ singlet with 
non-trivial $D'_3$ quantum number, ($\phi_a$), at the least we must add 
an $SO(10)$ singlet transforming as a ${\bf 1_B}$.

\section*{Appendix C. Results for charged fermion masses, mixing angles 
and neutrino oscillations}

In the paper~\cite{BRT} a global $\chi^2$ analysis has been
performed incorporating two (one) loop renormalization group[RG] running 
of dimensionless (dimensionful) parameters from $M_G$ to $M_Z$ in the MSSM,  
one loop radiative threshold corrections at $M_Z$, and 3 loop QCD (1 loop QED)
RG running below $M_Z$.  Electroweak symmetry breaking is obtained 
self-consistently from the effective potential at one loop, with all one 
loop threshold corrections included. This analysis is performed using the 
code of Bla\v zek et.al.~\cite{Blazek}.

In Table \ref{table:fit4nu} we give the 20 observables which enter the 
$\chi^2$ function, their experimental values and the uncertainty $\sigma$ 
(in parentheses). These are the results for one set of soft SUSY breaking 
parameters $m_0, \; M_{1/2}$ with all other parameters varied to obtain the 
best fit solution.  In most cases $\sigma$ is determined by the 1 standard 
deviation 
experimental uncertainty, however in some cases the theoretical uncertainty
($\sim$ 0.1\%) inherent in our renormalization group running and one
loop threshold corrections dominates.
\begin{table}
\caption[8]{
{\bf Charged fermion masses and mixing angles} \\  
Initial parameters: \\
 (1/$\alpha_G, \, M_G, \, \epsilon_3$) = ($24.52, \, 3.05 \cdot 10^{16}$
GeV,$\,
-4.08$\%), \\
 ($\lambda, \,$r$, \, \sigma, \, \epsilon, \, \rho, \, \epsilon^\prime$) =
($ 0.79, \,
12.4, \, 0.84, \, 0.011, \,  0.043,\,  0.0031$),\\
($\Phi_\sigma, \, \Phi_\epsilon, \, \Phi_\rho$) =  ($0.73, \, -1.21, \,
3.72$)rad, \\
($m_0, \, M_{1/2}, \, A_0, \, \mu(M_Z)$) = ($1000,\, 300, \, -1437, \,
110$) GeV,\\
($(m_{H_d}/m_0)^2, \, (m_{H_u}/m_0)^2, \, $tan$\beta$) = ($2.22,\, 1.65, \,
53.7$).
}
\label{table:fit4nu}
$$
\begin{array}{|l|c|l|}
\hline
{\rm Observable}  &{\rm Data}(\sigma) & Theory  \\
\mbox{ }   & {\rm (masses} & {\rm in\  \ GeV) }  \\
\hline   
\;\;\;M_Z            &  91.187 \ (0.091)  &  91.17          \\
\;\;\;M_W             &  80.388 \ (0.080)    &  80.40       \\
\;\;\;G_{\mu}\cdot 10^5   &  1.1664 \ (0.0012) &  1.166     \\
\;\;\;\alpha_{EM}^{-1} &  137.04 \ (0.14)  &  137.0         \\
\;\;\;\alpha_s(M_Z)    &  0.1190 \ (0.003)   &  0.1174       \\
\;\;\;\rho_{new}\cdot 10^3  & -1.20 \ (1.3) & +0.320   \\
\hline
\;\;\;M_t              &  173.8 \ (5.0)   &  175.0       \\
\;\;\;m_b(M_b)          &    4.260 \ (0.11)  &    4.328                  \\
\;\;\;M_b - M_c        &    3.400 \ (0.2)   &    3.421                 \\
\;\;\;m_s              &  0.180 \ (0.050)   &  0.148          \\
\;\;\;m_d/m_s          &  0.050 \ (0.015)   &  0.0589        \\
\;\;\;Q^{-2}           &  0.00203 \ (0.00020)  &  0.00201                \\
\;\;\;M_{\tau}         &  1.777 \ (0.0018)   &  1.776         \\
\;\;\;M_{\mu}          & 0.10566 \ (0.00011)   & .1057           \\
\;\;\;M_e \cdot 10^3      &  0.5110 \ (0.00051) &  0.5110  \\
 \;\;\;V_{us}         &  0.2205 \ (0.0026)      &  0.2205        \\
\;\;\;V_{cb}         & 0.03920 \ (0.0030)      &  0.0403           \\
\;\;\;V_{ub}/V_{cb}    &  0.0800 \ (0.02)    &  0.0691                 \\
\;\;\;\hat B_K          &  0.860 \ (0.08)    &  0.8703           \\
\hline
{B(b\!\rightarrow\! s \gamma)\!\cdot\!10^{4}}  &  3.000 \ (0.47) &
2.995  \\
\hline
  \multicolumn{2}{|l}{{\rm TOTAL}\;\;\;\; \chi^2}  3.39
            & \\
\hline
\end{array}
$$
\end{table}

For large tan$\beta$ there are 6 real Yukawa parameters and 3 complex 
phases $\Phi_\rho, \; \Phi_\epsilon$ and $\Phi_\sigma$.  With 13 fermion 
mass observables (charged fermion masses and mixing angles [$\hat{B}_K$ 
replacing $\epsilon_K$ as a ``measure of CP violation"]) 
we have 4 predictions.  For low tan$\beta$ , $\lambda \neq \xi$, we have 
one less prediction.  From Table \ref{table:fit4nu}  it is clear that this 
theory fits the low energy data quite well.

Finally, the squark, slepton, Higgs and gaugino spectrum of the theory is
consistent with all available data.  The lightest chargino and neutralino are
higgsino-like with the masses close to their respective experimental
limits. As an example of the additional predictions of this theory consider 
the CP violating mixing angles which may soon be observed at B factories.   
For the selected fit it was found
\begin{eqnarray}
(\sin 2\alpha, \, \sin 2\beta, \, \sin \gamma) = & (0.74, \, 0.54, \,
0.99)&
\end{eqnarray}  or equivalently the Wolfenstein parameters
\begin{eqnarray}
(\rho, \, \eta )    =      &( -0.04, \,      0.31)    &.
\end{eqnarray}

The results obtained in ref.~\cite{BRT} for the neutrino sector are presented 
in Tables ~\ref{t:4numass2} and ~\ref{t:4nuangles}.   The model has maximal 
$\nu_\mu \rightarrow \nu_\tau$ mixing to describe atmospheric neutrino data
and small mixing angle [SMA] $\nu_e \rightarrow \nu_s$ oscillations to fit 
solar neutrino data with SMA matter enhanced MSW oscillations.  The model 
{\em cannot} however fit the LSND $\nu_e \rightarrow \nu_\mu$ data.  

\protect
\begin{table}
\caption[3]{
{\bf Fit to atmospheric and solar neutrino \\ oscillations} \\
   \mbox{Initial parameters: ( 4 neutrinos \  with large tan$\beta$  ) }\ \
\ \

$m' = 7.11 \cdot 10^{-2}$ eV , \ $b$ = $-0.521$, \ $c$ = 0.278,\ $\Phi_b$ =
3.40rad
}
\label{t:4numass2}
$$
\begin{array}{|c|c|}
\hline
{\rm Observable} &{\rm Computed \;\; value} \\
\hline
\delta m^2_{atm}            &  3.2 \cdot 10^{-3} \ \rm eV^2          \\
\sin^2 2\theta_{atm}            &  1.08        \\
 \delta m^2_{sol}   &  4.2\cdot 10^{-6}  \ \rm eV^2  \\
\sin^2 2\theta_{sol} &  3.0\cdot 10^{-3}         \\
\hline
\end{array}$$
\end{table}

\protect
\begin{table}
\caption[3]{
{\bf Neutrino Masses and Mixings} \hspace{1.1cm} \\

\mbox{Mass eigenvalues [eV]: \ \  0.0, \ 0.002, \ 0.04, \ 0.07 \hspace{1cm}} \\
\mbox{Magnitude of neutrino mixing matrix  U$_{\alpha i}$ \hspace{1.7cm}}\\
\mbox{ $i = 1, \cdots, 4$ -- labels mass eigenstates. \hspace{1.5cm}} \\
\mbox{ $\alpha = \{ e, \ \mu, \ \tau, \ s \}$ labels flavor eigenstates.}
}
\label{t:4nuangles}
$$
\left[ \begin{array}{cccc}
0.998                   &  0.0204      & 0.0392   & 0.0529  \\
0.0689                  &    0.291     & 0.567    & 0.767  \\
0.317\cdot 10^{-3}      &  0.145       & 0.771    & 0.620  \\
0.284\cdot 10^{-3}      &   0.946      &  0.287     &  0.154 \\
\end{array} \right]$$
\end{table}

\end{document}